\documentclass[aps,prl,preprint,showpacs]{revtex4}

\usepackage{graphicx}

\begin{document}


\title{Delayed and Predictive Dynamics with Stochastic Time}


\author{Toru Ohira} 
\email[]{ohira@csl.sony.co.jp}
\homepage[]{www.csl.sony.co.jp/person/ohira}

\affiliation{Sony Computer Science Laboratories, Inc., Tokyo, Japan 
141-0022}


\date{\today}

\begin{abstract}
We present simple classical dynamical models to 
address the question of introducing a stochastic nature in a time variable. 
These models include noise in the time variable but not in the ``space" 
variable, which is opposite to the normal description of stochastic dynamics. 
The notable feature is that these models can induce a ``resonance" with 
varying noise strengths in the time variable. Thus, they provide a different 
mechanism for ``stochastic resonance'', which has been discussed within the 
normal context of stochastic dynamics. In a broader context, we expect these 
simple models and associated behaviours may serve as one of the starting 
points to construct and investigate an extended classical dynamical theory 
where both the space and time variables are taken as stochastic. We need to 
explore if such approaches can be further developed for both classical and 
quantum applications.
\end{abstract}


\maketitle


\section{Introduction}

``Time" is a concept that has gained a lot of attention from thinkers from 
virtually all disciplines\cite{davies1995}. In particular,  
our ordinary perception of space and time is not the same, and this gap has been 
appearing in a variety of contemplations of nature. It appears to be the main 
reason for the theory of relativity, which has conceptually brought space and 
time closer to receiving equal treatment, continues to fascinate and attract 
arguments from diverse fields. Also, issues such as ``directions" or 
''arrows" of time are current interests of research\cite{savitt1995}. Another 
manifestation of this gap is the treatment of noise or fluctuations in dealing 
with dynamical systems. When we consider dynamical systems, whether classical, quantum, or 
relativistic, time is commonly viewed as not having stochastic characteristics.
In stochastic dynamical theories, we associate noise 
and fluctuations with only ``space'' variables, such as the position of a 
particle, but not to the time variables. In quantum mechanics, the concept of 
time fluctuation is well accepted through the time-energy uncertainty 
principle. However, time is not treated as a dynamical quantum observable, 
and a clearer understanding has been explored\cite{Busch2002}.  

Against these backgrounds, our main theme of this paper is to consider 
fluctuation of time in classical dynamics through a presentation of simple 
models. We consider two cases. The first type includes fluctuations for the
time point of events to affect the dynamical change. The second type is such that
the fluctuation is in the flow of time variable itself. Both types exhibit
behaviors which are similar to stochastic resonance.

\section{Stochasticity in time of events}

The general differential equation form of the class of dynamics we 
discuss here is given as follows. 
\begin{equation}
{dx(t) \over dt} = f(x(\bar{t}),\bar{t}).
\end{equation}

Here, $x$ is the dynamical variable given as a function of time $t$, and $f$ is 
the "dynamical function" governing the dynamics. The difference from the 
normal dynamical equation appears in $\bar{t}$, which contains stochastic 
characteristics, and $t\neq\bar{t}$ in general. In other words, the change of 
$x(t)$ is governed by a function $f$, not from 
its "current" state $x(t)$, but its state at $\bar{t}$, which is given by some 
stochastic rules. We can define $\bar{t}$ in a variety of ways, as well as the 
function $f$ in normal dynamical equations. In the following, we will present 
two models and investigate their properties through computer simulations. 
In order to avoid ambiguity and for simplicity, these models are dynamical 
map systems incorporating the basic ideas of the general definition given 
above.
The first model we start with is a special case, which is given as follows.
\begin{eqnarray}
x_{n+1} & = & x_{n} + r \cos(\bar{t_{n}}),\\
\bar{t_{n}} & = & \Delta t (n +\sigma \xi).
\end{eqnarray}
Here, $x_n$ is the dynamical variable and $\bar{t_{n}}$ is the time 
containing noise term $\xi$. The dynamics progress by 
incrementing $n$ with constant parameters, $r$, $\Delta t$, and $\sigma$. 
We fix the noise $\xi$ to take a value between $(-1, 1)$ with a uniform 
probability. This is a special case as the dynamical function $f = r 
\cos(t)$ depends only on time and not on $x_n$. When $\sigma=0$, it 
recovers ordinary dynamics with a sinusoidal path. We have performed 
computer simulations for this dynamical map with different sets of constants. 
The examples of the time series and associated power spectrum are shown in 
Fig. 1. The most notable characteristics are as we change the noise 
``strength'' or ``width'' given by $\sigma$, a rhythmic behavior appears and 
disappears. This is shown by the change in the peaks of the power spectrums, 
an example of which we plotted in Figure 1f. The spacing of the peaks 
appearing in Figure 1f is one half the period of the dynamical function, 
indicating $\sigma$ is a source for producing in and out of phase dynamics. 
At the same time, we note that the role of noise is important, as when we set 
$\xi$ to be a constant rather than a random variable, the resulting 
dynamics is a simple sinusoidal time series and a change in the peak height 
is not observed. The periodical oscillatory dynamics emerging by ``tuning of 
noise'' has been studied under the name of ``stochastic 
resonance''\cite{wiesenfeld-moss95,bulsara96,gammaitoni98} and 
investigated in a variety of 
fields\cite{mcnamara88,longtin-moss91,collins1995,chapeau2003,lee2003}. 
We observe here a similar behavior with stochasticity in time rather than in
space variable.
\begin{figure}
  \includegraphics[width=.5\textwidth]{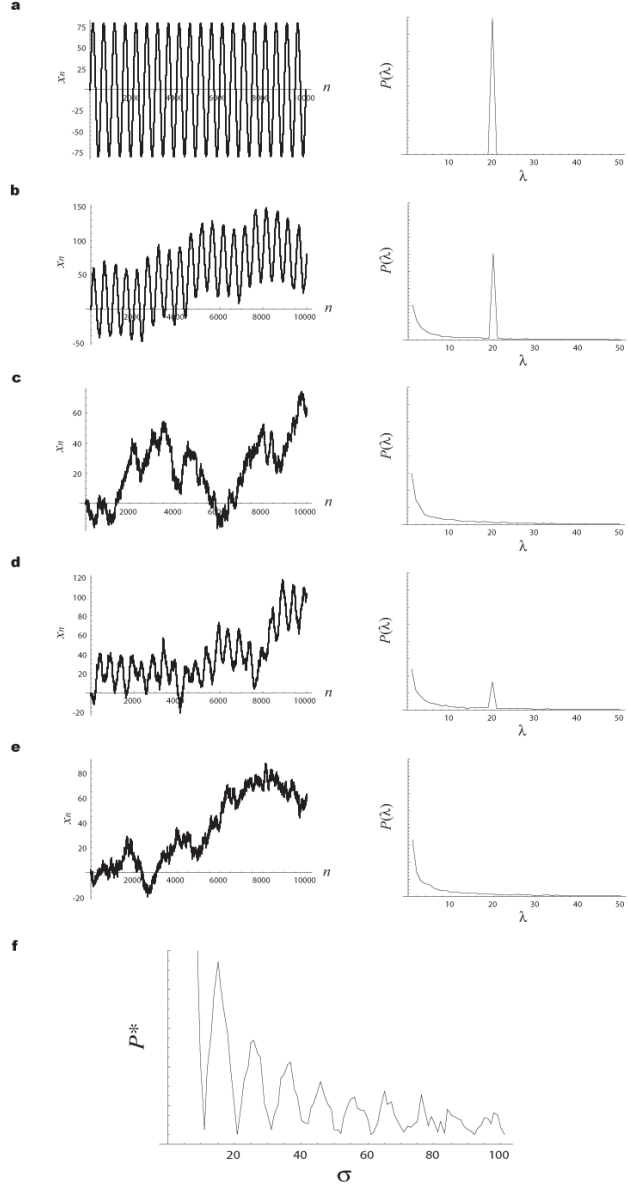}
  \caption{Dynamics and power spectrum of sinusoidal function model with 
stochastic event time. This is an example of the dynamics and associated power 
spectrum through the simulation of the model given in Eq. (2). The 
parameters are set as $r = 1$, $\Delta t = 0.004 \pi $, and the noise widths 
$\sigma$ are set to (a) $\sigma=0$, (b) $\sigma=125$, (c) $\sigma=250$, 
(d) $\sigma=375$, and (e) $\sigma=500$. The simulation is performed up to 
$L=10000$ steps and 20 averages are taken for the power spectrum. The unit of frequency $\lambda$ is set as ${1 \over L}$, 
and the power $P(\lambda)$ is in arbitrary units. Graph (f) is the peak 
height $P^{*}$ as a function of the noise width $\sigma$.}
 \end{figure}

Our second model is given as follows, which again shows a resonance-type 
behavior with a tuning of noise. 
\begin{eqnarray}
x_{n+1} & = & (1-b) x_{n} + c \theta [\bar{x}(\bar{t_{n}})],\\
\bar{t_{n}} & = & \phi (n +\sigma \xi).
\end{eqnarray}
Here, $b$ and $c$ are the parameters, and $\phi (z)$ is a function that gives 
the closest integer to $z$. In this equation, $\xi$ is the same noise term as in 
the first model and its width is controlled by $\sigma$, and $\theta (x)$ is a 
``threshold function" (Figure 2(a)) such that,
\begin{equation}
\theta (x) = \left\{ \begin{array}{ll}
                        -1 & x>0 \\
                        0  & x=0 \\
                        1  & x<0 \\
                        \end{array}
                \right. 
\end{equation}
In this model, we further need to define $\bar{x}(\bar{t_{n}})$, which is 
given as follows.
\begin{equation}
\bar{x}(\bar{t_{n}}) = \left\{ \begin{array}{cc}
                        x_0 & \bar{t_{n}} \leq 0 \\
                        x_{\bar{t_{n}}}  & 0 < \bar{t_{n}} \leq n \\
                        x_{n} + ({\bar{t_{n}}}-n)(x_n - x_{n-1})   & n < 
\bar{t_{n}} \\
                        \end{array}
                \right.
\end{equation}

\begin{figure}
  \includegraphics[width=.7\textwidth]{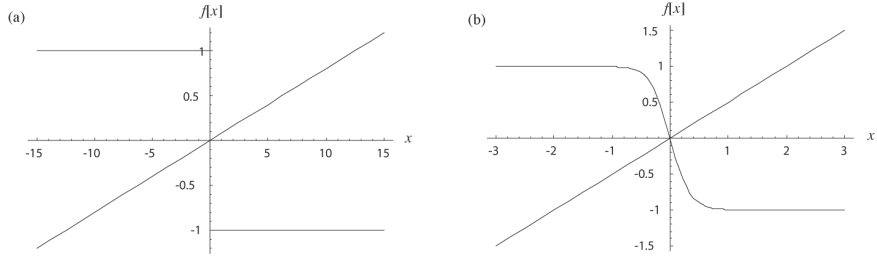}
  \caption{Dynamical functions $f(x)$ with parameters as examples of simulations presented in this paper. (a) Threshold
function. 
The Straight line has slope of $b = 0.08$. (b) Negative feedback function with parameters $\beta = 6$. Straight line has 
slope of $\alpha = 0.5$.}
 \end{figure}

The meaning of this definition is that when $\bar{t_{n}}$ is in the past 
($\bar{t_{n}}\leq n$), the value of $x$ at that past point is used. One the 
other hand, if $\bar{t_{n}}$ is in the future ($n < \bar{t_{n}}$), we estimate 
$x$ as the value that would be obtained if the same rate of current change 
extends to the time duration from the present to the future point. 
Qualitatively, this method of projecting into the future is one of the most 
commonly used for estimating  population and national debt, for instance. 
The same estimation is used for the recent study in ``predictive dynamical 
systems''\cite{ohira06}.

The computer simulation of this model is performed with various parameter 
sets, examples of which are shown in Fig. 3. Without noise 
($\sigma=0$), the basic dynamics are decaying to the origin, $x=0$. With 
the introduction of noise with an increasing width, we begin to observe more 
rhythmic behaviors. This is seen with the peaks of the power spectrum. The 
signal to noise ratio, calculated as the ratio of the peak to the background 
height level of the power spectrum, is the main characterization of the 
stochastic resonance.
 
\begin{figure}
  \includegraphics[width=.5\textwidth]{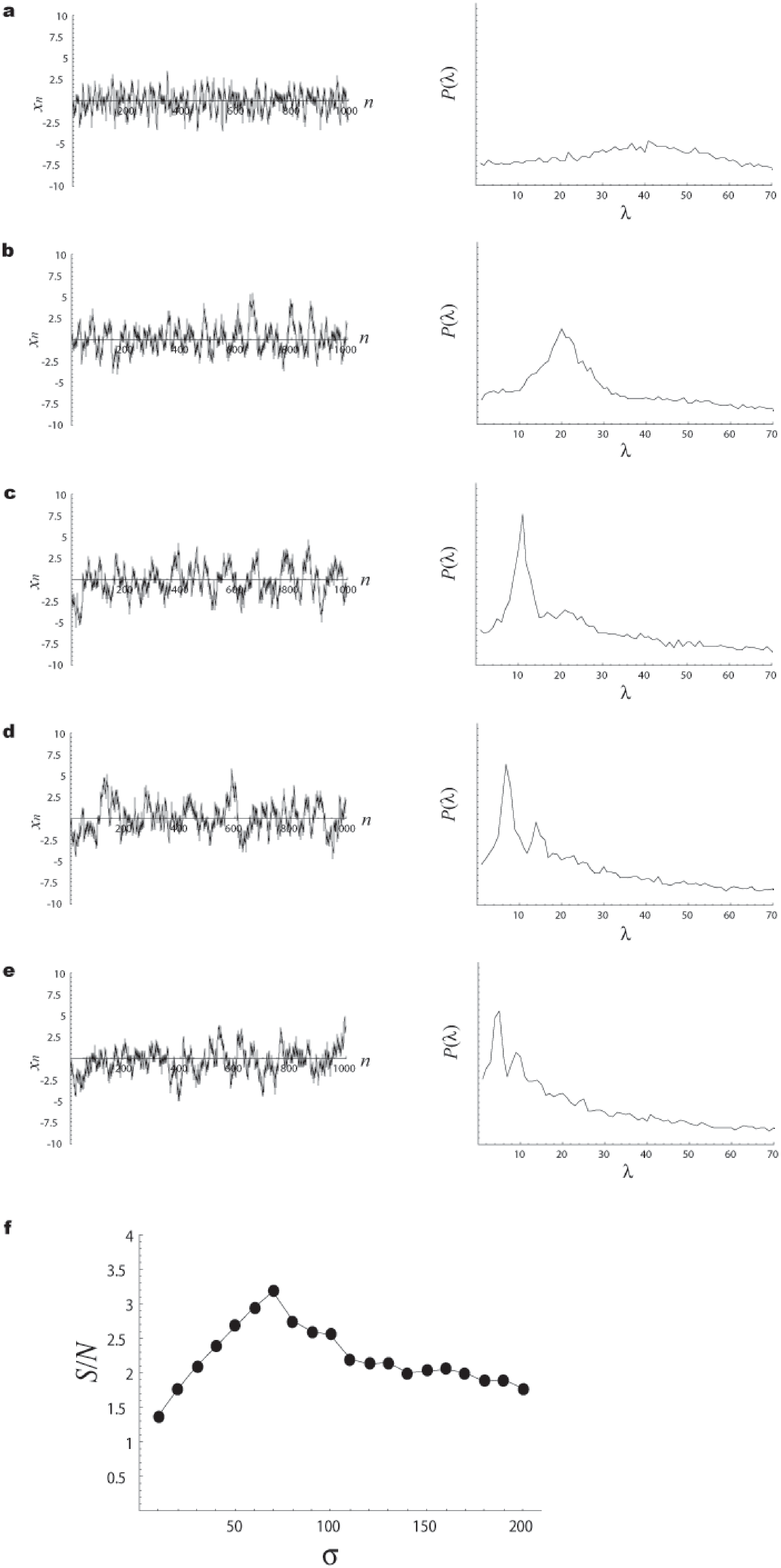}
  \caption{Dynamics and power spectrum of threshold function model with 
stochastic event time. This is an example of the dynamics and associated power 
spectrum through the simulation of the model given in Eq. (3).  The 
parameters are set as $b = 0.08$, $c=1.0$, and the noise widths 
$\sigma$ are set to (a) $\sigma=10$, (b) $\sigma=30$, (c) $\sigma=70$, (d) 
$\sigma=120$, and (e) $\sigma=200$. The simulation is performed up to 
$L=10000$ steps 20 averages are taken for the power spectrum. The unit of frequency $\lambda$ is set as ${1 \over L}$, 
and the power $P(\lambda)$ is in arbitrary units. Graph (f) is the signal to 
noise ratio ${S \over N}$ at the peak height as a function of the noise width 
$\sigma$.}
 \end{figure}

If we plot this as a function of the noise strength, we 
can obtain the resonance curve shown in Figure 3f. This indicates that this 
signal to noise ratio goes through maximum at the optimal noise level with 
other parameters fixed. Therefore, once again, the time fluctuation is acting 
in a constructive way.

We would now like to discuss a couple of points with these models. First of all, 
it should be noted that, in these models, the basic flow and/or ``arrow'' of 
time is the same as in ordinary dynamical equations. In other words, the 
rate of change of the dynamical variable $x_{n}$ is dictated by stochastically 
chosen point $\bar{t_{n}}$ on the time axis, but the ordering of $x_{n}$ is 
kept by $n$. From another point of view, these models can be considered a 
stochastic extension of the combined 
delayed\cite{mackey77,cooke82,milton89,ohira-yamane00,frank-beek01} 
and predictive\cite{ohira06} dynamical systems. Thus, the stochastic nature 
of time is only partially implemented here.

\section{Stochasticity in the time flow}

We now turn our attention to the second type of models, in which the noise
is in the time flow itself. There are several ways to achieve this effect.
The discussion here is limited with a stochastic time flow combined with
a delayed dynamics. The general form of the dynamical equation 
is given as follows.
\begin{equation}
{dx(\bar{t}) \over d\bar{t}} = f(x(t), x(t-\tau)),
\end{equation}
where $\bar{t}$ is again a stochastic variable. It should be noted that
we now have a stochasticity in time flow. In particular, we focus on the
following corresponding map.
\begin{eqnarray}
x_{n+\xi} & = & f(x_{n}, x(n-\tau)).
\end{eqnarray}
Here, $\xi$ is the stochastic variable which can take either $+1$ or $-1$
with certain probabilities. Thus even though the dynamics progress by iteration
of $n$, it occasionally ``goes back in time'' with the occurrence of $\xi=-1$.
Let the probability of $\xi=-1$ be given by $p$. Then naturally, with $p=0$,
this map reduces to a normal delayed map. The dynamical function is
chosen to be a negative feedback function (Figure 2(b)) and the concrete map model we
study is given as follows.
\begin{eqnarray}
x_{n+\xi} & = & x_{n} + d\delta (-\alpha x_{n} - {2 \over {1+e^{-\beta x_{n-\tau}}}} -1).
\end{eqnarray}

When there is no noise in time flow, this map has the origin as the stable fixed point
with no delay. The linear stability analysis gives the critical delay $\tau_c$, at which the stability of the fixed point is
lost, as $0.59 \sim   \tau_c d\delta$. The larger delay gives an oscillatory dynamical path.
We have found, through computer simulations, that an interesting behavior
arises when delay is smaller than this critical delay. The tuned noise in the time
flow gives the system more tendency for oscillatory behavior. In other words,
adjusting the value of $p$ controlling $\xi$ induces oscillatory behaviors.
Some examples are shown in Figure 4. With increasing probability for a time
flow to reverse, i.e., with $p$ increasing, we observe oscillatory behaviors
as shown both in the sample dynamical path as well as in the corresponding power spectrum.
However, with $p$ increased enough, the oscillatory behaviors begins to 
deteriorate. The change of the peak heights is shown in Figure 5. 
Again, we see a phenomena which resembles stochastic resonance. 
A resonance with delay and noise, called
``delayed stochastic resonance''\cite{ohira-sato}, has been proposed with an additive noise in
``space''. Analytical understanding 
of the mechanism of our model here is yet to be explored. However,
this mechanism with stochastic time flow is clearly of different type and new. 

\begin{figure}
  \includegraphics[width=.8\textwidth]{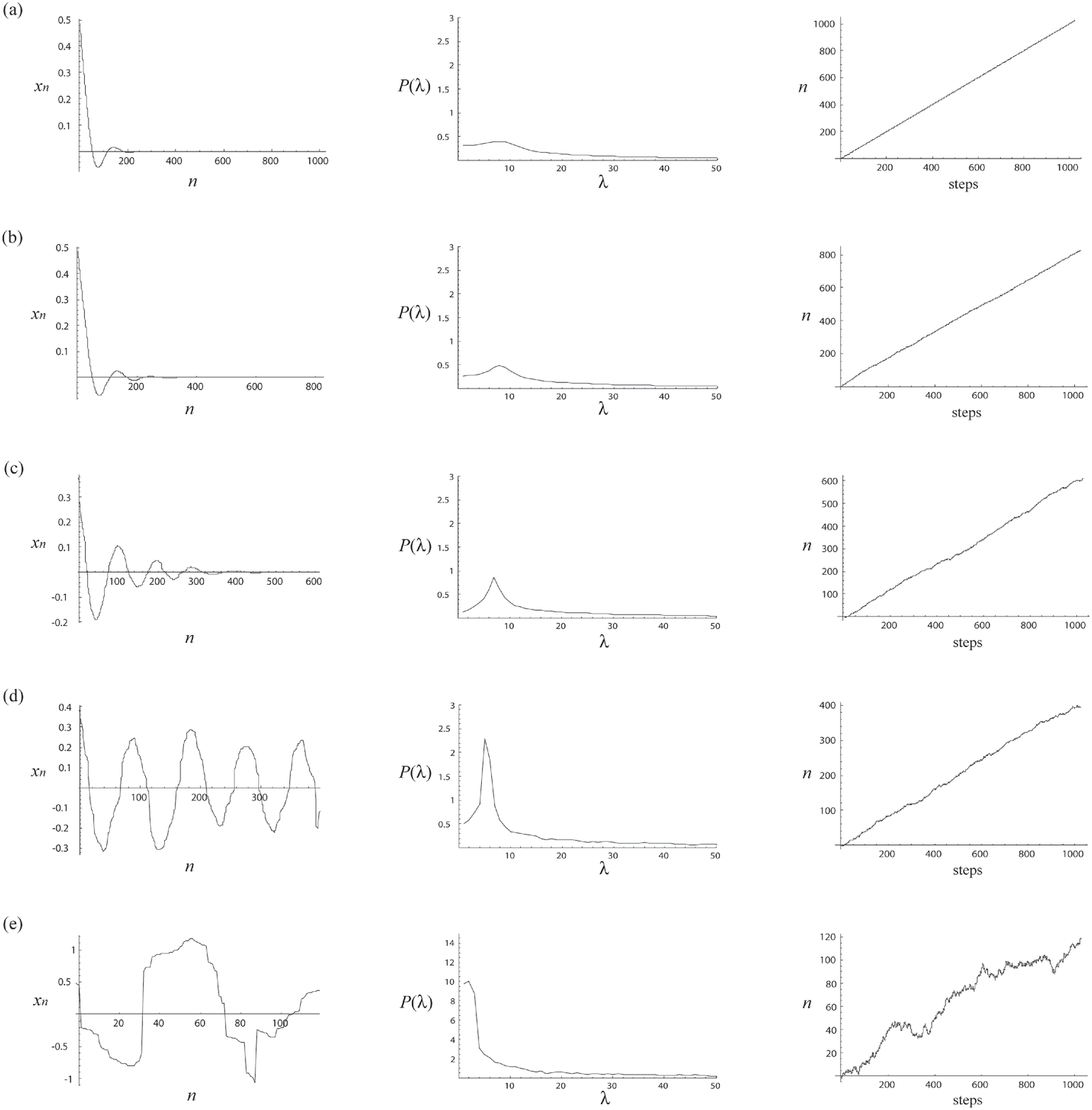}
  \caption{Dynamics (left) and power spectrum (middle) of delayed dynamical model with 
stochastic time flow (right). This is an example of the dynamics and associated power 
spectrum through the simulation of the model given in Eq. (10) with the probability 
$p$ of stochastic time flow varied.  The 
parameters are set as $\alpha = 0.5$, $\beta = 6$, $d\delta=0.01$, $\tau=25$ and the 
stochastic time flow parameter $p$ are set to (a) $p=0$, (b) $p=0.1$, (c) $p=0.2$, (d) 
$p=0.3$, and (e) $p=0.45$. The simulation is performed up to 
$L=1000$ steps and 20 averages are taken for the power spectrum. The unit of frequency $\lambda$ is set as ${1 \over L}$, 
and the power $P(\lambda)$ is in arbitrary units. }
 \end{figure}

\begin{figure}
\includegraphics[width=.55\textwidth]{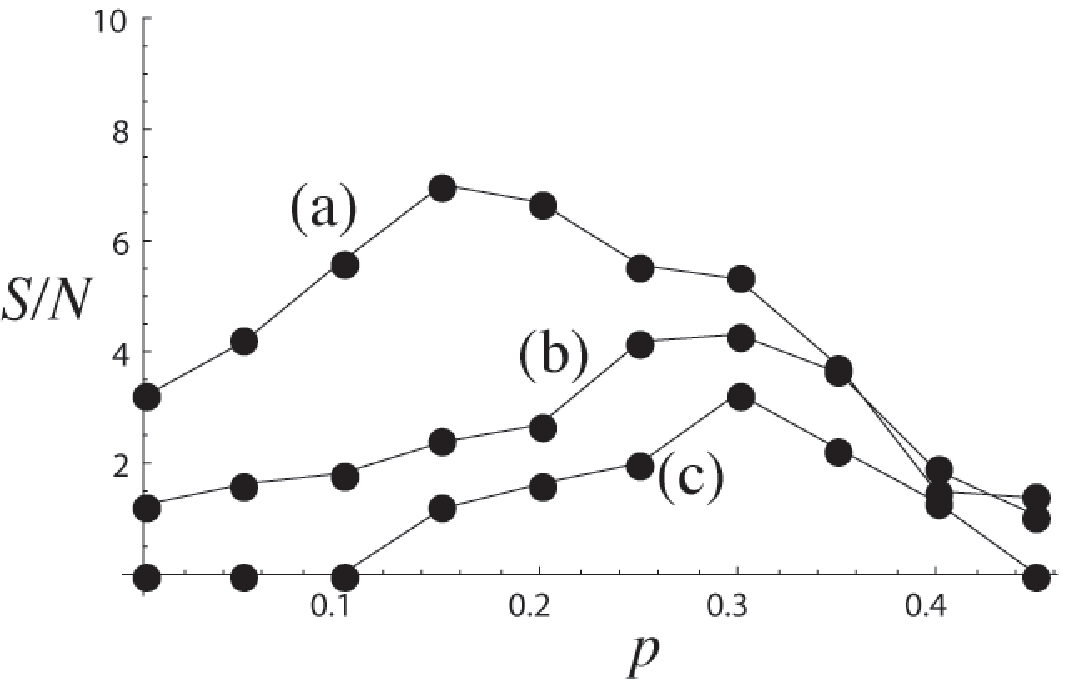}
\caption{
The signal to 
noise ratio ${S \over N}$ at the peak height as a function of the probability  
$p$ of stochastic time flow. The parameter settings are the same as in the Figure 4 with
the delay varied as (a) $\tau=40$, (b) $\tau=25$ and (c) $\tau=15$.
}
 \end{figure}

\section{Discussion}

We would like to now discuss a couple of points with respect to these models.
First, we can extend these models to include fluctuations in the variable 
$x_{n}$. Proceeding in this way, we have a picture of dynamical systems with 
fluctuations on both the time and space axes. The analytical frameworks and 
tools for such descriptions need to be developed along with a search for 
appropriate applications. For an example of appropriate applications, we 
may think of modeling a stick balancing on a human fingertip. Recent 
experiments have found that most of the corrective motion observed has 
shorter time scales compared to the human reaction 
time\cite{cabrera-milton02,cabrera-milton04B,cabrera-etal04}. This may be 
the result of intricate mixtures of physiological delays, predictions, and 
physical fluctuations. Models incorporating special fluctuations and delays 
have been considered, but bringing in the effect from both the past and 
future through the stochastic time may help to further develop it. 

Another 
direction may be to extend the path integral formalisms along this approach. 
Whether this type of extension bridges into quantum pictures and/or leads to 
an alternative understanding of such properties as time-energy uncertainty 
relations requires further investigation.

Finally, if these models are capturing some aspects of reality, particularly with respect to
stochasticity in time flow, this resonance 
may be used as an experimental indication for probing fluctuations or 
stochasticity in time. We have previously proposed ``delayed stochastic 
resonance''\cite{ohira-sato}, a resonance by the interplay of noise and delay. 
It was theoretically extended\cite{tsimring01}, and recently the effect was 
experimentally observed through a solid-sate laser system with a feedback 
loop\cite{masoller}. It is left for the future to see if an analogous 
experimental test could be developed with respect to stochastic time.

\end{document}